\documentclass[12pt]{article}

\usepackage{amsmath,amssymb}

\global\arraycolsep=1pt
\newcommand{\bel}[1]{\begin{equation}\label{#1}}                     
\newcommand{\bal}[1]{\begin{eqnarray}\label{#1}}                     
\newcommand{\be}{\begin{equation}}
\newcommand{\ee}{\end{equation}}

\newcommand{\de}{\mathrm{d}}

\newcommand{\scr}{\scriptstyle}
\newcommand{\qq}{\qquad}

\renewcommand{\thefootnote}{\fnsymbol{footnote}}

\begin{document}

\begin{flushright}
August, 2007 \\
OCU-PHYS 273 \\
\end{flushright}
\vspace{20mm}

\begin{center}
{\bf\Large
Closed conformal Killing-Yano tensor\\
and Kerr-NUT-de Sitter spacetime uniqueness
}
\end{center}

\begin{center}

\vspace{15mm}

Tsuyoshi Houri$^a$\footnote{
\texttt{houri@sci.osaka-cu.ac.jp}
}, 
Takeshi Oota$^b$\footnote{
\texttt{toota@sci.osaka-cu.ac.jp}
} and
Yukinori Yasui$^a$\footnote{
\texttt{yasui@sci.osaka-cu.ac.jp}
}

\vspace{10mm}

${}^a$
\textit{
Department of Mathematics and Physics, Graduate School of Science,\\
Osaka City University\\
3-3-138 Sugimoto, Sumiyoshi,
Osaka 558-8585, JAPAN
}
\vspace{5mm}

\textit{
${{}^b}$
Osaka City University
Advanced Mathematical Institute (OCAMI)\\
3-3-138 Sugimoto, Sumiyoshi,
Osaka 558-8585, JAPAN
}

\vspace{5mm}

\end{center}
\vspace{8mm}

%%%%%%%%%%%%%%%%%%%%%%%%%%%%%%%%%%%%%%%%%%%%%%%%%%%%%%%%%%%%%%%%%%%%%%%%%%%%%%%%%%%%
\begin{abstract}
We study spacetimes with
a closed conformal Killing-Yano tensor.
It is  shown that the $D$-dimensional Kerr-NUT-de Sitter spacetime constructed
by Chen-L\"{u}-Pope is the only spacetime admitting a rank-$2$
closed conformal Killing-Yano tensor with a certain symmetry.
\end{abstract}
%%%%%%%%%%%%%%%%%%%%%%%%%%%%%%%%%%%%%%%%%%%%%%%%%%%%%%%%%%%%%%%%%%%%%%%%%%%%%%%%%%%%%

\vspace{25mm}

\newpage

\renewcommand{\thefootnote}{\arabic{footnote}}
\setcounter{footnote}{0}

%%%%%%%%%%%%%%%%%%%%%%%%%%%%%%%%%%%%%%%%%%%%%%%%%%%%%%%%%%%%%%%%%%%%%%%%%%%%%%%%%%%%%%%%%%

Higher dimensional black hole solutions have attracted 
renewed interests in the recent developments
of supergravity and superstring theories. Recently, 
the $D$-dimensional 
Kerr-NUT-de Sitter metrics were constructed by \cite{CLP}. 
All the known vacuum type D black hole solutions are included in these metrics \cite{HHOY}. 
Kerr-NUT-de Sitter metrics are also interesting from the point of view of AdS/CFT
correspondence. Indeed, odd-dimensional metrics lead to Sasaki-Einstein metrics
by taking  BPS limit \cite{CLP,HSY,CLPP1,CLPP2}
and even-dimensional metrics lead to Calabi-Yau metrics
in the limit \cite{CLP,OY,LP}. Especially, the five-dimensional Sasaki-Einstein metrics have
emerged quite naturally in the AdS/CFT correspondence.

On the other hand, it has been shown that geodesic motion in the
Kerr-NUT-de Sitter spacetime is integrable for all dimensions \cite{FK,KF,PKVK,FKK,KKPF,KKPV}. 
Indeed, the constants of motion that are in involution
can be explicitly constructed from a rank-$2$ closed conformal
Killing-Yano (CKY) tensor. In this paper, using 
a geometric characterisation of the separation of variables in the 
Hamilton-Jacobi equation \cite{HOY},  we study spacetimes
with
a rank-$2$ closed CKY tensor.

The rank-$2$ CKY tensor is defined as a two-form 
\be
h = \frac{1}{2} h_{ab}\, \de x^a \wedge \de x^b, \qq
h_{ab} = - h_{ba}
\ee
satisfying the equation \cite{tac}
\begin{equation} \label{CKY}
\nabla_a h_{bc}+\nabla_b h_{ac}= 2 \xi_c g_{ab}- \xi_a g_{bc}-\xi_b g_{ac}.
\end{equation} 
The vector field $\xi_a$ is called
the associated vector of $h_{ab}$, which is given by
\be
\xi_a = \frac{1}{D-1} \nabla^b h_{ba}.
\ee

\noindent
{\bf{Theorem}}
\textit{Let us assume the existence of a single rank-$2$  CKY tensor $h$
for $D$-dimensional spacetime $(M,g)$ satisfying the conditions,
\begin{equation}
(a)~dh=0,~~~
(b)~\mathcal{L}_{\xi} g=0,~~~
(c)~\mathcal{L}_{\xi} h=0.
\end{equation}
Then, $M$ is only the Kerr-NUT-de Sitter spacetime}\footnote{We require a 
further technical condition which will be detailed in the proof.
See the assumption below eq.\eqref{MC2}.}.\\

The Kerr-NUT-de Sitter metric takes the form \cite{CLP}:\\
\noindent
(a)~$D=2n$
\begin{equation} \label{eveng}
g=\sum_{\mu=1}^{n}\frac{dx_{\mu}^2}{Q_{\mu}}+\sum_{\mu=1}^{n} Q_{\mu}
\left( \sum_{j=0}^{n-1} A^{(j)}_{\mu} d\psi^j \right)^2.
\end{equation}
\noindent
(b)~$D=2n+1$
\begin{equation} \label{oddg}
g=\sum_{\mu=1}^{n}\frac{dx_{\mu}^2}{Q_{\mu}}+\sum_{\mu=1}^{n} Q_{\mu}
\left( \sum_{j=0}^{n-1} A^{(j)}_{\mu} d\psi^j \right)^2+ S
\left( \sum_{j=0}^{n} A^{(j)} d\psi^j \right)^2.
\end{equation}
The functions $Q_{\mu}$ are given by
\begin{equation}
Q_{\mu}=\frac{X_{\mu}}{U_{\mu}}, \qq
U_{\mu}=\prod_{\stackrel{\scr \nu=1}{(\nu \ne \mu)}}^n (x_{\mu}^2-x_{\nu}^2),
\end{equation}
where $X_{\mu}$ is an arbitrary function depending only on $x_{\mu}$\,\footnote{
We call the metric Kerr-NUT-de Sitter for an arbitrary $X_{\mu}$. The existence of $h$ does not restrict
the metric to be Einstein.} and
\begin{equation}
A^{(k)}_{\mu}=\sum_{\stackrel{\scr 1 \le \nu_1 < \cdots < \nu_k \le n}{ (\nu_i \ne \mu)}} 
x_{\nu_1}^2x_{\nu_2}^2 \cdots x_{\nu_k}^2, ~~
A^{(k)}=\sum_{1 \le \nu_1 < \cdots < \nu_k \le n } x_{\nu_1}^2x_{\nu_2}^2
\cdots x_{\nu_k}^2,
\end{equation}
$(A_{\mu}^{(0)}=A^{(0)}=1)$ and $S=c/A^{(n)}$ with a constant $c$.\\

In the following we briefly describe the proof (see \cite{HOY2} for detailed analysis).  
The wedge product of two CKY
tensors is again a CKY tensor and so the wedge powers $h^{(j)}= h \wedge \cdots \wedge h$
are CKY tensors. 
The condition $(a)$ means that the Hodge dual $(D-2j)$-forms $f^{(j)}= \ast h^{(j)}$ 
are Killing-Yano tensors:
\begin{equation}
\nabla_{(a_1} f^{(j)}_{a_2)a_3 \cdots a_{D-2j+1}}=0.
\end{equation}
These Killing-Yano tensors generate the rank-$2$ Killing tensors $K^{(j)}$
obeying the equation $
\nabla_{(a}K^{(j)}_{bc)}=0$~. 
Under the condition $(a)$ the Killing tensors $K^{(j)}$ are mutually
commuting \cite{KKPF,KKPV},
\be 
[K^{(i)},~ K^{(j)}]_S =0. 
\ee
The bracket $[~~ , ~~]_S$ represents a symmetric Schouten product. The equation can be
written as
\begin{equation}
K^{(i)}_{d(a} \nabla^{d} K^{(j)}_{bc)}-K^{(j)}_{d(a} \nabla^{d} K^{(i)}_{bc)}=0.
\end{equation}
Let us define the vector fields $\eta^{(j)}$ by
\begin{equation}
\eta^{(j)}_a = K^{(j)~b}_{~~a}\xi_b. 
\end{equation}
Then we have
\begin{equation}
\nabla_{(a}\eta^{(j)}_{b)} =
\frac{1}{2}\mathcal{L}_{\xi}K^{(j)}_{ab}-\nabla_{\xi}K^{(j)}_{ab},
\end{equation}
which vanishes by the conditions (b) and (c),~i.e. $\eta^{(j)}$ are Killing vectors.
We can show that Killing vectors $\eta^{(i)}$ and Killing tensors $K^{(j)}$ 
are mutually commuting \cite{HOY},
\bel{MC2} 
[\eta^{(i)},~ K^{(j)}]_S =0,~~[\eta^{(i)},~ \eta^{(j)}] =0. 
\ee
Here, we assume that the Killing tensors $K^{(j)}$ and
$K^{(ij)}=\eta^{(i)} \otimes \eta^{(j)}+\eta^{(j)} \otimes \eta^{(i)}$ 
are independent. Therefore all the separability conditions of the geodesic Hamilton-Jacobi
equation are satisfied \cite{HOY}.\\

Let $y^a$ be geodesic separable coordinates of $D=(n+k)$-dimensional spacetime $M$:
\begin{equation}
y^{a}=(x^{\mu},~\psi^i), ~~~\mu=1,2, \cdots ,n,~~~i=0,1, \cdots, k-1,
\end{equation}
where $k=n$ $(k=n+1)$ for $D$ even (odd). In these coordinates
the commuting Killing vectors $\eta^{(j)}~ (j=0,1, \cdots , k-1)$ are written as
$\eta^{(j)}=\partial/\partial \psi^j$. 
From \cite{BF,KM,ben} the inverse
metric components are of the form,
\begin{equation} \label{metric}
g^{\mu \mu}=\bar{\phi}^{\mu}_{(0)}(x),
~~g^{ij}=\sum_{\mu=1}^{n} \zeta^{ij}_{\mu} (x^{\mu})\bar{\phi}^{\mu}_{(0)}(x),
\end{equation}
and the components of the Killing tensors $K^{(j)}$ are given by
\begin{equation}
K^{(j)\mu\nu}=\delta^{\mu\nu}\bar{\phi}^{\mu}_{(j)}(x),~~K^{(j)\mu i}=0,
~~K^{(j)i \ell}
=\sum_{\mu=1}^{n} \zeta^{i \ell}_{\mu}(x^{\mu})\bar{\phi}^{\mu}_{(j)}(x).
\end{equation}
Here, $\bar{\phi}^{\mu}_{(j)}$ is the $j$-th column of the inverse of an $n \times n$ 
St\"{a}ckel matrix $(\phi^{(j)}_{\mu})$,~i.e. 
each element depends on the variable corresponding to
the lower index only: $\phi^{(j)}_{\mu}(x^{\mu})$.
It should be noticed that the Killing tensors are constructed from CKY tensors, so that
they obey the following recursion relations as linear operators \cite{HOY}:
\begin{equation}
K^{(j)}=A^{(j)} I - Q K^{(j-1)},
\end{equation}
where $I$ is an identity operator and $Q$ is defined by
\begin{equation}
Q^{a}{}_{b}=-h^{a}{}_{c} h^{c}{}_{b}.
\end{equation}
Here $A^{(j)}$ is given by
\be
\det\!{}^{1/2}( I + \beta Q) = \sum_{j=0}^n A^{(j)} \beta^j.
\ee
Note that the equation \eqref{CKY} with the condition $(a)$ is equivalent to
\bel{CCKY}
\nabla_a h_{bc} = \xi_c g_{ab} - \xi_b g_{ac}.
\ee
We can further restrict the unknown functions $\bar{\phi}^{\mu}_{(0)}$ 
and $\zeta^{ij}_{\mu}$
in the metric \eqref{metric}. 
This is analyzed by considering the equation \eqref{CCKY} with $\xi = \eta^{(0)}$, 
and finally we find the Kerr-NUT-de Sitter metric \eqref{eveng} or \eqref{oddg}.

As a crosscheck of our theorem, we confirmed by the direct calculation
that a CKY tensor satisfying $(a)$, $(b)$ and $(c)$ does not
exist in the five-dimensional black ring background \cite{ER}.

\vspace{5mm}

%%%%%%%%%%%%%%%%%%%%%%%%%%%%%%%%%%%%%%%%
\noindent
{\bf{Acknowledgements}}

\vspace{3mm}

We would like to thank Yoshitake Hashimoto and Ryushi Goto for discussions.
This work is supported by the 21 COE program
``Construction of wide-angle mathematical basis focused on knots".
The work of YY is supported by the Grant-in Aid for Scientific
Research (No. 19540304 and No. 19540098)
from Japan Ministry of Education. 
The work of TO is supported by the Grant-in Aid for Scientific
Research (No. 18540285 and No. 19540304)
from Japan Ministry of Education.

%%%%%%%%%%%%%%%%%%%%%%%%%%%%%%%%%%%%%%%%%%%%%%%%%%%%%%%%%%%%%%%%%%%%%%%%%%%%%%%%%%%%%%%%%%%%%%%%%%%%%%%%%%%%%%%%%%%5

\end{document}